\documentclass[jmp,amsmath,amssymb,preprint]{revtex4-1}

\usepackage{aeguill}
\usepackage{graphicx}
\usepackage{dcolumn}
\usepackage{bm}

\usepackage{url}
\usepackage{verbatim}

\draft 

\begin{document}


\title{Suppression of the thermal hysteresis in magnetocaloric MnAs thin film by highly charged ion bombardment} 



\author{M. Trassinelli}
\email[]{martino.trassinelli@insp.jussieu.fr}
\author{M.~Marangolo}
\author{M.~Eddrief}
\author{V.H.~Etgens}
\altaffiliation[Also at ]{Universit\'e de Versailles Saint-Quentin en Yvelines, 55, Avenue de Paris - 78035 Versailles, France.}
\author{V.~Gafton}
\author{S.~Hidki}
\author{E.~Lacaze}
\author{E.~Lamour}
\author{C.~Prigent}
\author{J.-P.~Rozet}
\author{S.~Steydli}
\author{Y.~Zheng}
\author{D.~Vernhet}
\affiliation{CNRS, UMR 7588, Institut des NanoSciences de Paris (INSP), F-75005, Paris, France}
\affiliation{Sorbonne Universit\'es, UPMC Univ Paris 06, UMR 7588, INSP, F-75005, Paris, France}

\date{\today}

\begin{abstract}
We present the investigation on the modifications of structural and magnetic properties of MnAs thin film epitaxially grown on GaAs induced by slow highly charged ions bombardment under well-controlled conditions. 
The ion-induced defects facilitate the nucleation of one phase with respect to the other in the first-order magneto-structural MnAs transition
with a consequent suppression of thermal hysteresis without any significant perturbation on the other structural and magnetic properties. 
In particular, the irradiated film keeps the giant magnetocaloric effect at room temperature opening new perspective on magnetic refrigeration technology for everyday use.

\end{abstract}


\maketitle 

At present, the application of the magnetocaloric effect (MCE) as an alternative method for refrigeration is one of the great technological challenges. 
Compared to the common gas-compression/expansion method, MCE has an higher efficiency with absence of moving parts, and a consequently small environmental impact and maintenance.
Materials showing conventional MCE are characterized by a second-order magnetic transition.
In giant-MCE (GMCE) materials, a magneto-structural first-order transition generally occurs.
The search of materials with a GMCE close to room temperature is of great interest and it is mainly obtained by varying material composition\cite{Pecharsky1997a,Pecharsky1997,Wada2001,deCampos2006,Rocco2007,Sun2008,Cui2009} or, more recently, by applying an external strain to bulk\cite{Liu2012} or to thin films.\cite{Mosca2008,Duquesne2012,Moya2013}
However, first-order transitions exhibit  a considerable thermal hysteresis $\Delta T_\text{hys}$, which makes GMCE materials difficult to handle in applications for real refrigerators that work cyclically.
Much effort has been made for reducing this hysteresis. 
In the past years, this reduction has been obtained by doping bulk manganese arsenide (MnAs)\cite{Rocco2007,Sun2008,Cui2009}, where a suppression of the thermal hysteresis has been reached but only for low intensity magnetic field\cite{Sun2008,Cui2009} ($H=0.01$~T).

Another interesting way to change the magnetic properties of thin films is the bombardment and implantation of ions.
Nevertheless, up to present, only monocharged ions have been used to irradiate materials exhibiting a second-order transition exclusively. \cite{Chappert1998,Kaminsky2001,Zhang2003,Zhang2004,Muller2005,Fassbender2006,Cook2011}  

Here, we investigate the modifications of MnAs thin film epitaxially grown on a GaAs substrate submitted to the bombardment of highly charged ions.
MnAs is one of the more promising GMCE materials.
It exhibits a large change of magnetic entropy (typically\cite{Wada2001,Wada2005} 
$\Delta S(T=\text{cst})\approx - 30$~J\ Kg$^{-1}$ K$^{-1}$
 for a field change $\Delta H=2$~T) in proximity of its transition close to room temperature ($T_C = 313$~K)
corresponding to a large refrigeration power (that depends on the $\Delta$S integral over a temperature interval) up to 200~J Kg$^{-1}$.
This ferromagnetic--non-ferromagnetic transition is associated with the magneto-structural phase transition from hexagonal ($\alpha$-phase, NiAs-type) to orthorhombic ($\beta$-phase, MnP-type).
Compared to bulk materials, in MnAs thin films the strain of the substrate disturbs the phase transition that leads to the $\alpha - \beta$ phase coexistence.
This is characterized by a self-organization with longitudinal alternating regions over a large range of temperatures (290--320~K), generating a consequent modification of the magnetic properties of the film.\cite{Daweritz2006}
In particular, the phase coexistence reduces the maximum value of $\Delta$S(T) but keeps the same refrigeration power. Indeed the $\Delta$S per mole of material portion passing from one phase to another is still very high, which characterizes the giant MCE materials.\cite{Mosca2008}
The period $\lambda$ of the self-organization depends linearly on the MnAs film thickness $t$ with the relationship\cite{Kaganer2002,Kastner2002}  $\lambda \approx 4.8\ t$.

MnAs epilayers investigated here are grown by molecular beam epitaxy (MBE) on GaAs(001) substrate. 
The deposed MnAs is oriented with the $\alpha$-MnAs$[0001]$ and $\beta$-MnAs$[001]$ axis parallel to GaAs$[\bar110]$.
At the end of the growth process, $150 \pm 10$~nm thick samples are capped in situ with an amorphous As layer in order to prevent the MnAs oxidation before the ion bombardment. 
More details on the growth process can be found in Ref.~\citenum{Breitwieser2009}.

The ion irradiation is performed at the SIMPA facility\cite{Gumberidze2010} (French acronym for highly charged ion source of Paris) that includes an electron-cyclotron resonance ion source coupled to a dedicated ultra-high vacuum beam line. 
The different samples,
 with a surface of about $4 \times 5$~mm$^2$ obtained from the same wafer,
 are irradiated with a beam of Ne$^{9+}$ ions with a kinetic energy of 90~keV (4.5~keV/u).
The incidence angle between the ion beam and the sample surface is set at 60$^\circ$, for having an average penetration depth of the ions corresponding to the half-thickness of the MnAs film\cite{Ziegler2008} with a consequent maximization effect of ion irradiation.\cite{Zhang2004}
The ion--sample collision zone is continuously monitored with a visible-light sensitive CCD camera and a X-ray solid-state detector during the irradiation.
Only a negligible fraction of ions is deposited in the GaAs substrate excluding the possibility of MnAs-GaAs mixing.\cite{Fassbender2006}
Different ion beam bombardment durations at 0.5~$\mu$A beam intensity, from 5 to several thousands of seconds, and corresponding to a fluence between $\Phi = 1.6\times10^{12}$ and  $1.6\times10^{15}$~ions/cm$^2$, are applied on different samples coming from the same growth. 
The potential energy of the ions, which depends on their charge state, contributes marginally, with only 3.1~keV,  making the dependency on the ion charge insignificant in the bombardment.
More details about the irradiation process can be found in Ref.~\citenum{Trassinelli2013}.

\begin{figure}
\includegraphics[clip=true,trim= 40 35 50 50, width=0.7\columnwidth]{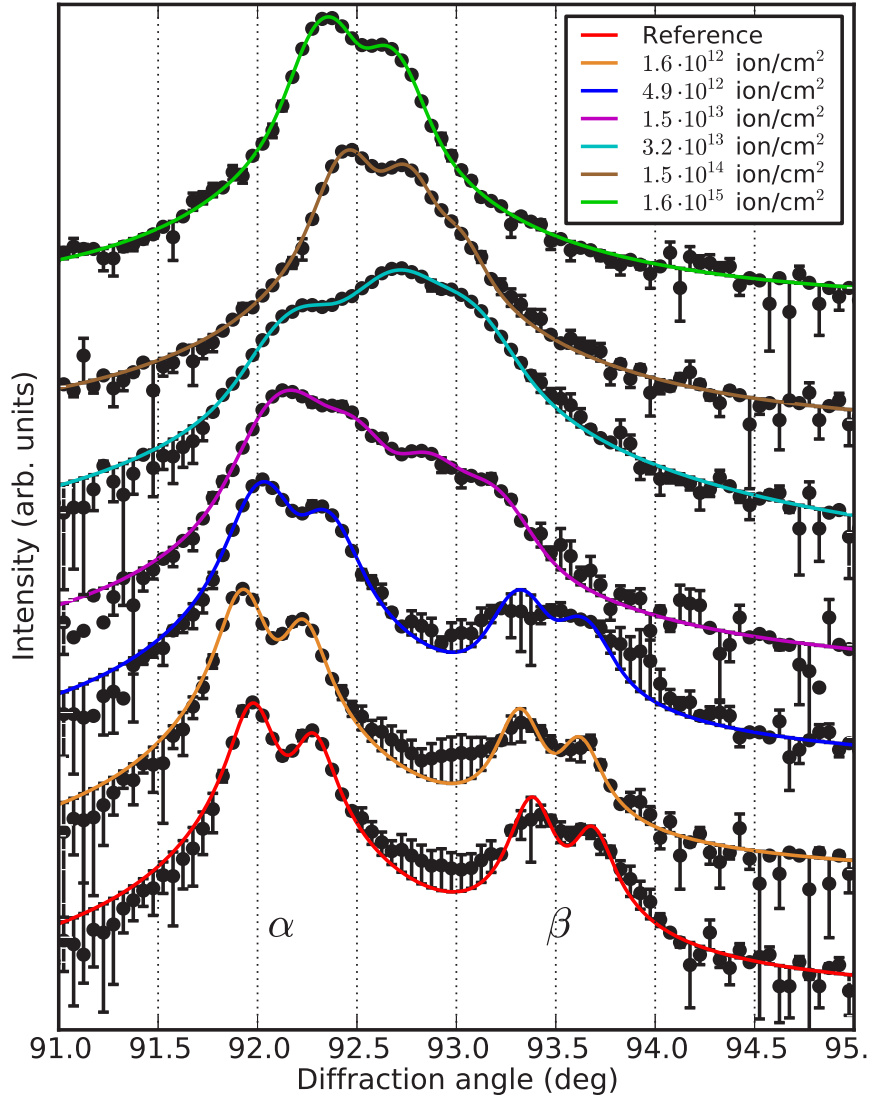}
\caption{\label{fig:XRD} $\alpha$-MnAs$(\bar3300)$ (left) and $\beta$-MnAs$(060)$ (right) reflection peaks of samples submitted to different ion fluences $\Phi$. The double peak structure is due to the CuK$\alpha_{1,2}$ emission used in the diffractometer. 
The full lines represent the result of the profile fits.}
\end{figure}

\begin{figure}
\includegraphics[clip=true,trim= 100 90 100 90,width=\columnwidth]{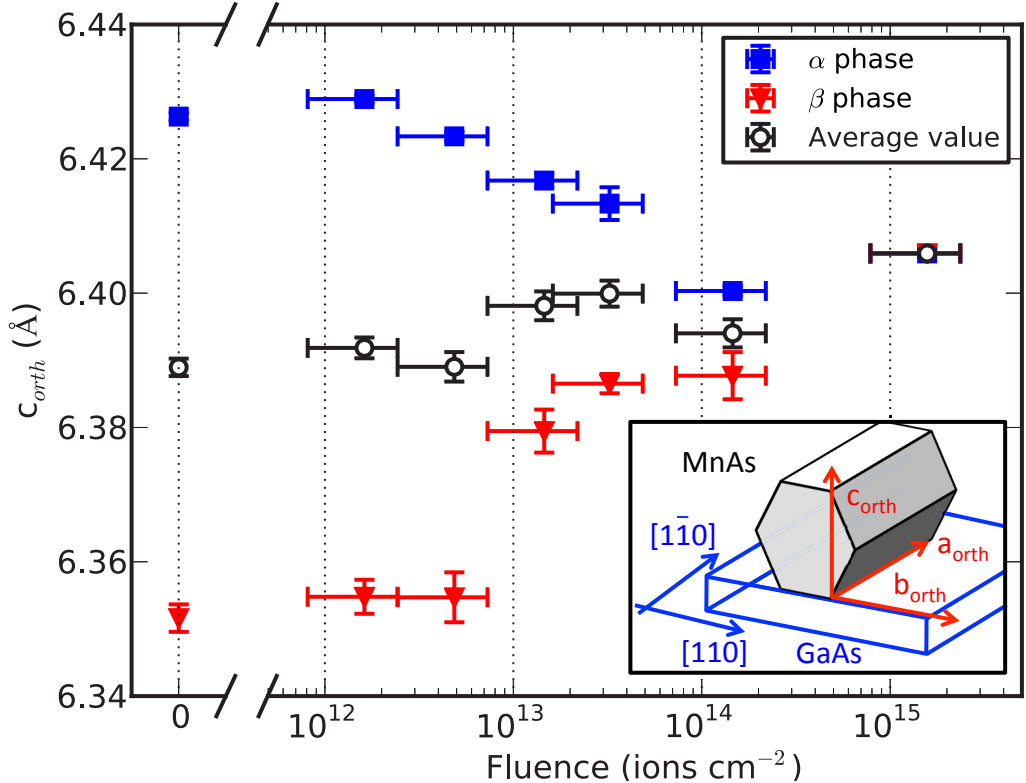}
\caption{\label{fig:d-spacing} (Color online) Plot of the $c_\text{orth}$ values of the $\alpha$-  and $\beta$-phases extracted from the diffraction curves as a function of the ion fluence $\Phi$.
In the inset, scheme of the MnAs film with its orientation relative to the GaAs(001) substrate.}
\end{figure}

After highly charged ions impact, sample properties modification and their dependency on the ion fluence are studied using different techniques, namely: X-ray diffraction (XRD), magnetic force microscopy (MFM) and sample magnetometry (with a vibrating sample magnetometer, VSM, and a superconducting quantum interference device, SQUID magnetometer).
With the XRD (model PANalytical XPert MRD), structural changes are investigated at room temperature ($T =  293\pm1$~K). 
From X-ray reflectivity, the MnAs layer thickness of the different samples is evaluated to the constant value of 150~nm, whatever the ion fluence, demonstrating that sputtering effects are negligible.
XRD measurements at Bragg angles are used to determine the $\alpha$-  and $\beta$-phase crystal spacing as a function of the fluence $\Phi$ 
and are presented in  Fig.~\ref{fig:XRD}. 
From this detailed analysis, the $\alpha$-MnAs$(\bar3300)$ and $\beta$-MnAs$(060)$ reflection peaks are clearly identified for the non-irradiated sample (curve in the bottom), the $\beta$ peaks being much less intense at 293~K. 
At low fluence, the $\alpha$  and $\beta$ diffraction reflections are well separated. 
At the highest fluence, they merge resulting to an unique diffraction reflection.
From the angle difference between the MnAs and the substrate GaAs reflections, the lattice constants  of the two phases are measured. 
We use here the orthorhombic crystallographic system (shown in the inset of Fig.~\ref{fig:d-spacing}), which is more appropriate since the residual strain breaks off the hexagonal symmetry of the $\alpha$-phase.
The lattice values of the axis perpendicular to the surface, $c_\text{orth}$, are presented in Fig.~\ref{fig:d-spacing}.
For the non-irradiated sample, $c_\text{orth}$ values are comparable to the literature\cite{Adriano2006} values for similar sample thickness at 293~K: $c_\text{orth}(\alpha)=6.44$~\AA  \ and $c_\text{orth}(\beta)=6.37$~\AA.
Differently to the preliminary data survey presented in Ref.~\citenum{Trassinelli2013}, we can observe in Fig.~\ref{fig:d-spacing} that the contribution from the two phases is still clearly distinguishable for any value of the fluence except for the highest.
 The presence of two distinguished structural phases indicates that the associated magnetic transition remains of first-order type.
 $c_\text{orth}(\alpha)$ continuously decreases when increasing $\Phi$, whilst $c_\text{orth}(\beta)$ increases until the merging of the two diffraction peaks.
 This progressive bridging suggests an increasing of the strain between the zones of different phases due to their spatial fragmentation, in analogy with the theoretical results presented in Ref.~\citenum{Kaganer2002}.

\begin{figure}
\includegraphics[clip=true,trim= 70 170 70 50,width=0.8\columnwidth]{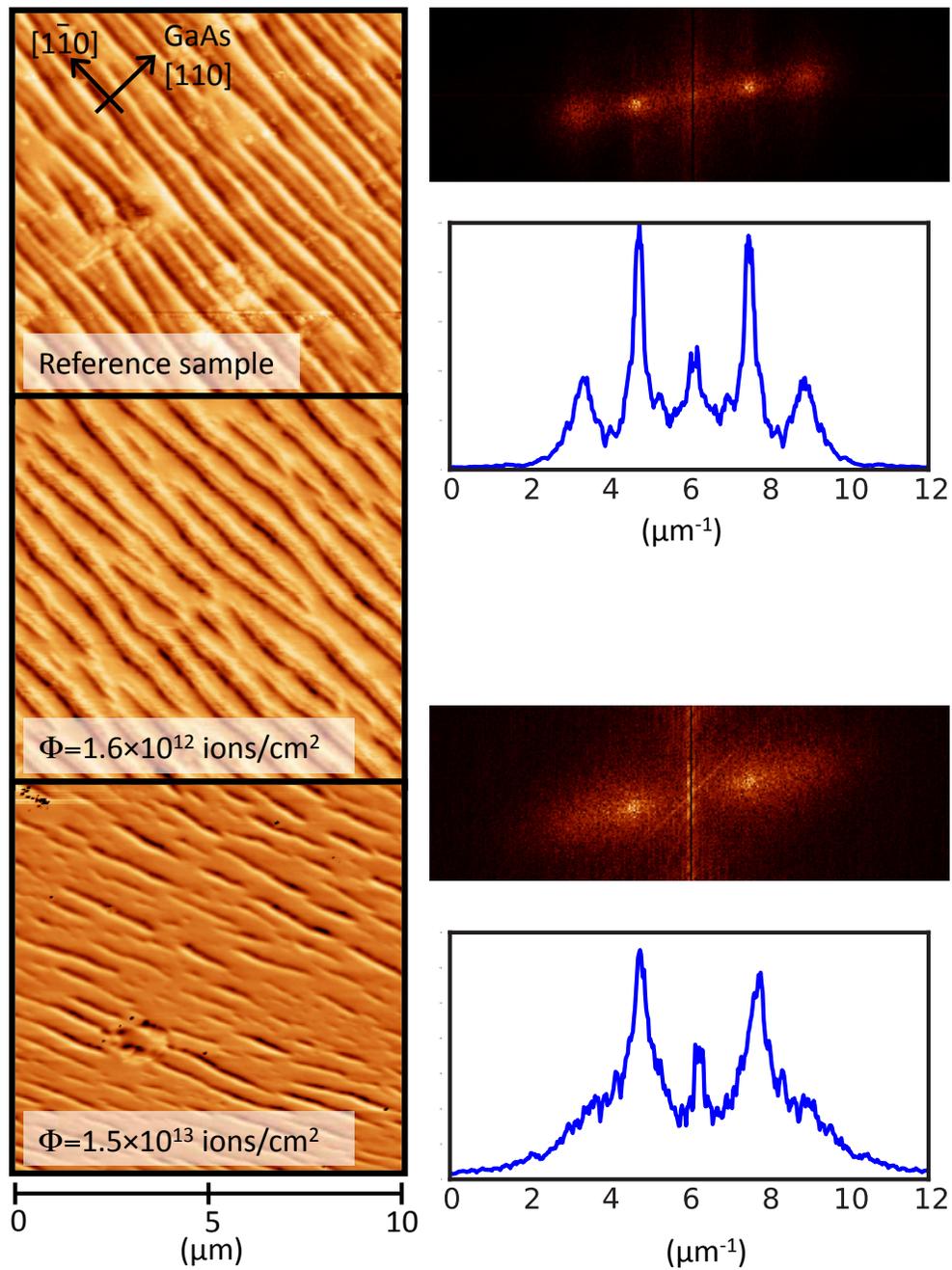}
\caption{\label{fig:MFM} (Color online) On the left, MFM images of the reference sample (top) and of samples submitted to different ion fluences. The regular spacing due to the GaAs substrate constrains is visible in all images. In irradiated samples, random defects due to the ion bombardment introduce a distortion of the regular pattern.
On the right, Fourier transforms and intensity profiles are presented for the reference and the most irradiate samples.
}
\end{figure}

A direct observation of the $\alpha$ - and $\beta$-phase zones layout can also be obtained with a MFM (Bruker Multimode AFM microscope equipped with a magnetic tip coated with Co/Cr, model MESP).
For this survey performed at 293~K, a well-defined procedure has been applied before each measurement.
The samples are demagnetized at higher temperature ($T \approx 340$ ~K) and then magnetized along the surface parallel to the easy axis $b_\text{orth}$.
The contrast detected by the MFM corresponds to the out-of-plane component of the stray magnetic field emanating from $\alpha$ stripes.
The resulting images are presented in Fig.~\ref{fig:MFM}. For the non-irradiated sample (top), the regular arrangement between $\alpha$- and $\beta$-phase is well visible.
The Fourier transform of a large sample area ($40 \times 40~\mu$m) shows clearly the first and the second order maxima corresponding to $\lambda = 0.73~ \mu$m, close to the expected periodicity. 
MFM images of the bombarded samples show that the stripe-type structure is more and more distorted, but not suppressed.
The Fourier transform of the irradiated sample with $1.5 \times 10^{13}$~ions/cm$^2$ presents indeed maxima at the same position than for the non-irradiated sample,
with simply a larger dispersion,
due to the random defects produced by ion impact on the regular pattern.
Quantitatively, an increase of the full width at half maximum from $0.31 \pm 0.01~\mu$m$^{-1}$ to $0.83 \pm 0.08~\mu$m$^{-1}$ is measured.
We recall that the main periodicity $\lambda$ is intrinsically related to the structural difference between the  $\alpha$- and $\beta$-phases,\cite{Kaganer2002} and no extra phase of MnAs has to be invoked.\cite{Trassinelli2013}
The observed  increasing fragmentation of the $\alpha-\beta$ regions and the absence of modification in the $\lambda$ periodicity confirm the interpretation made above for the XRD data.
For $\Phi >1.5 \times 10^{13}$~ions/cm$^2$, magnetic imaging becomes impossible due to the low out-of-plane magnetic field (stripes disappearance). 

Even if the structural properties are not strongly modified by the ion bombardment, the presence of additional seeding defects might perturb the phase transition and then modify the transition temperature, the thermal hysteresis and the giant magnetocaloric properties.
These aspects are investigated by measuring the magnetic moment, the coercivity and the magnetic anisotropy of the samples at different temperatures and magnetic fields. 
Information about the coercivity field, anisotropy and, more generally, magnetic hysteresis cycle are extracted with a vibrating sample magnetometer (VSM, in the Quantum Design PPMS 9T) from  $(M(H),H)$ hysteresis curves of the magnetization $M$ at different temperatures $T$ with a variable magnetic field $H$. 
After a depolarization at 350~K and $H=0$, each sample has been brought at a defined temperature $T$. Then, the magnetic moment is recorded continuously during the field $H$ variation between +1 and -1~T.
\begin{figure}
\includegraphics[width=0.9\columnwidth]{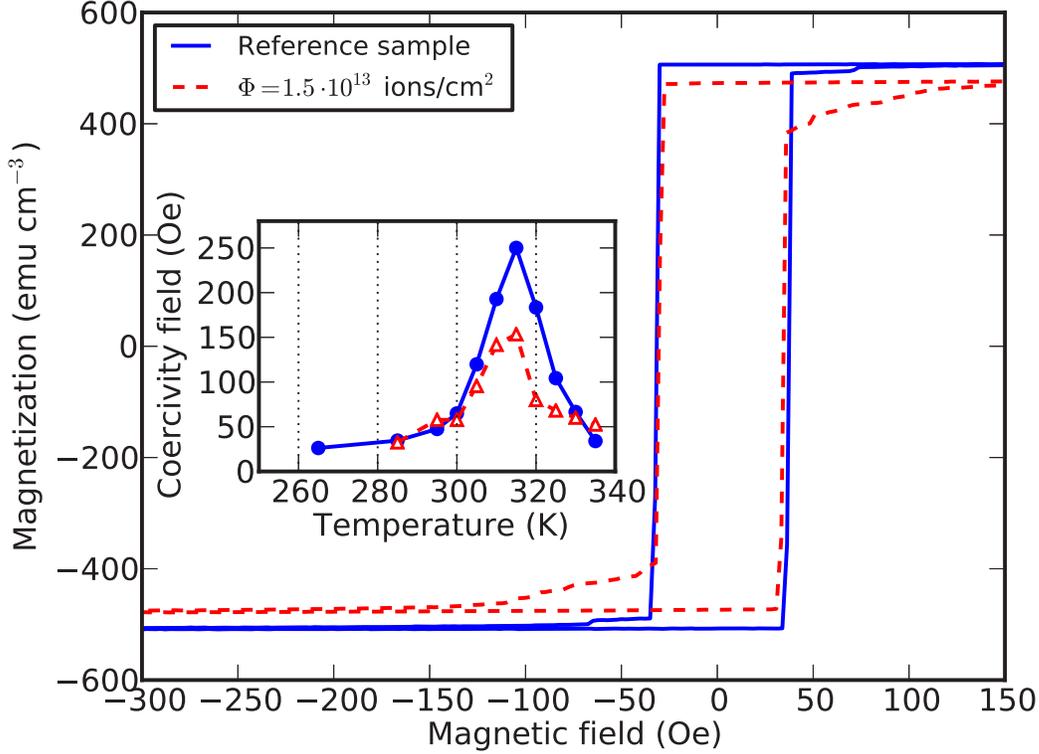}
\caption{\label{fig:MH} (Color online) Magnetization as a function of external magnetic field at the temperature of 285~K for the reference and for the irradiated samples.
In the inset, the dependency of the coercivity field on the temperature is shown.}
\end{figure}
No large difference on the hysteresis cycles $(M(H),H)$ is noticeable between the reference and the irradiated sample (see Fig.~\ref{fig:MH}). The presence of the ion-induced defects produces additional pinning on the mobility of the magnetic domains, visible by the presence of ``wings'' on the magnetic cycle. 
In contrast, the nucleation of the magnetic domains is unchanged.
The temperature dependency of the associated coercivity field (inset in Fig.~\ref{fig:MH}) shows the characteristic peak on the $\alpha - \beta$ coexisting zone.\cite{Steren2006}
A small reduction of its maximum value is noticeable for the bombarded sample, from 250~Oe to 150-200~Oe, but a constant value of about 30~Oe is found below $\sim300$~K, independently on the ion fluence.

\begin{figure}
\includegraphics[width=\columnwidth]{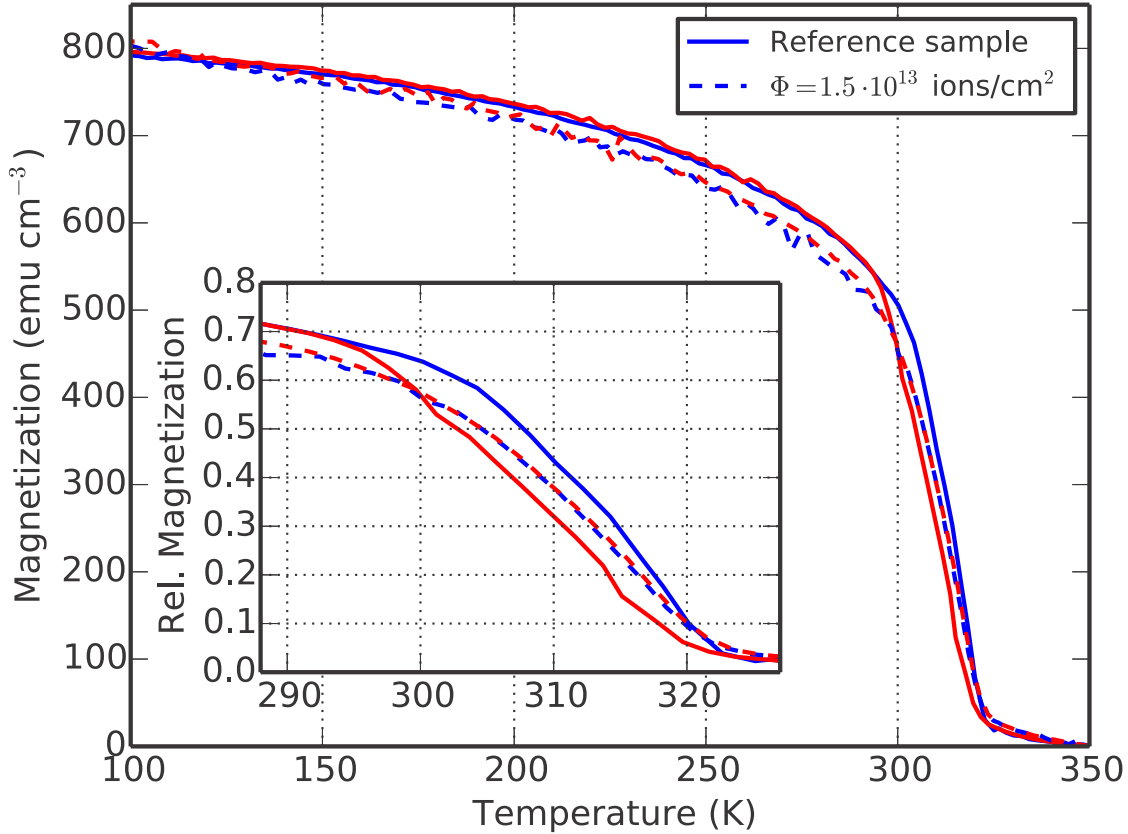}
\caption{\label{fig:MT} (Color online) Magnetization as a function of temperature for the reference (solid lines) and for the irradiated samples (dashed lines). Data obtained by a temperature increase (from colder temperatures) and decrease (from hotter temperatures) are presented in blue and red, respectively.
In the inset, the magnetization relative to saturation close to the transition region is shown.}
\end{figure}

The samples magnetization dependency on the temperature, with a fixed $H$ applied, is obtained with a SQUID magnetometer (Quantum Design MPMS-XL).
The measurement procedure is the following: 
i) each sample is initially brought to 350~K with $H=0$; 
ii) the sample is cooled down to 100~K and then a magnetic field $H=1$~T is applied;
iii) the magnetic moment is recorded continuously during the temperature variation from 100 to 350~K, and then back to 100~K, with a sweep rate of $\pm2$~K/min.
The results are presented in Fig.~\ref{fig:MT}, where $(M(T),T)$ curves corresponding to an irradiated and the reference samples are shown.
%
At low temperature, the saturation magnetization values $M_\text{sat}$ of the reference and irradiated samples are comparable within the experimental uncertainty of 1--2\%.
Similarly, the transition temperature $T_C$, defined here as the temperature for which $M(T_C) = M_\text{sat}/2$, is for both samples around 305~K. 
In contrast, differently from the reference sample, characterized by $\Delta T_\text{hys} \approx 5$~K for $H=1$~T, the thermal hysteresis disappears in the irradiated samples.

After the observation of the suppression of $\Delta T_\text{hys}$, it is interesting to check wether the magnetocaloric properties of the irradiated samples have been also modified or not. 
The MCE is evaluated from the dependence of the magnetization at different temperatures and external fields in the SQUID magnetometer, following a procedure similar to  that described in Ref.~\citenum{Mosca2008}:
i) each sample is initially brought to 350~K with $H=0$; 
ii) a magnetic field $H$ (with a starting value equal to 2~T) is applied;
iii) the sample is cooled down to 150~K;
iv) the magnetic moment is recorded continuously during the temperature variation from 150 to 350~K with a sweep rate of +2~K/min;
v) at $T=350$~K, the magnetic field is decreased with a step of 0.2~T;
and then again from ii) to v) is repeated until $H=0$.
At a given temperature $T$, the magnetic entropy change is calculated numerically by\cite{GschneidnerJr2005}
$\Delta S(T,\Delta H) = \int_{H_i}^{H_f} (\partial M/ \partial T)_H \ dH$,
for a magnetic field variation $\Delta H=H_f - H_i$.
\begin{figure}
\includegraphics[clip=true,trim= 0 5 0 15,width=0.9\columnwidth]{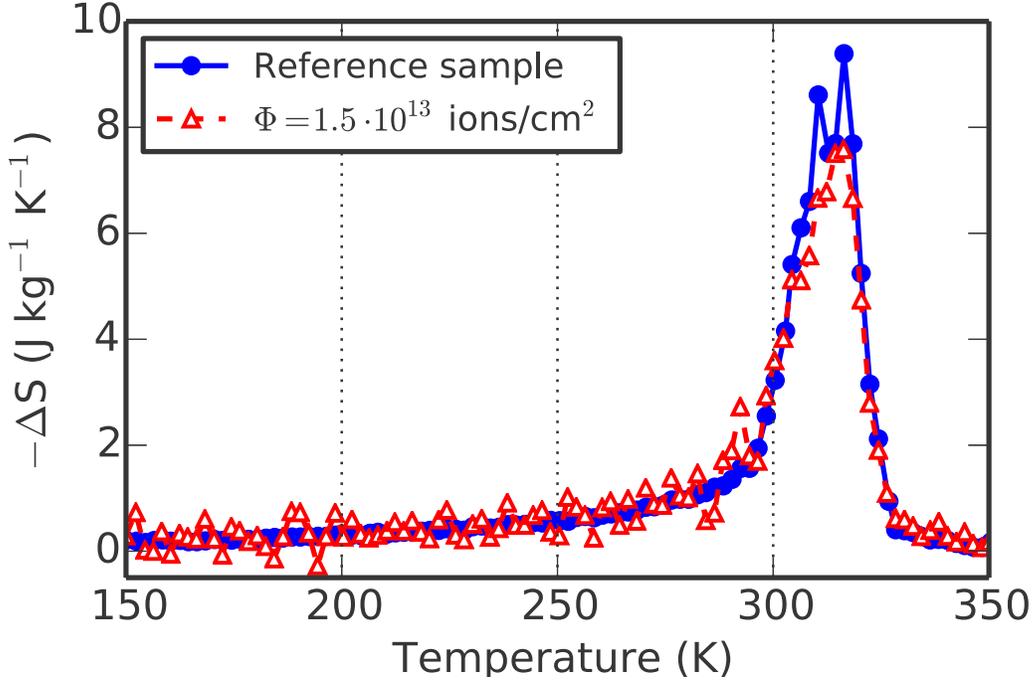}
\caption{\label{fig:MCE} (Color online) Magnetic entropy change as a function of temperature determined from magnetization data of the reference and irradiated sample for a magnetic field change from 0 to 2~T.}
\end{figure}
As presented in Fig.~\ref{fig:MCE}, the magnetic entropy change is only weakly affected by the ion bombardment, the integrated value of $\Delta S$ between 290 and 330~K decreases only from 
175 to 163~J/kg ($-7\%$). 
The irradiated MnAs thin film results in keeping the giant magnetocaloric properties at room temperature accompanied by a fully reversible behavior in $M(T)$ curves.

Summarizing the different observations, we can conclude that the main effect of the highly charged ion bombardment on MnAs thin films is the disappearance of the thermal hysteresis occurring in the magneto-structural phase transition.
The defects induced by the ion collision facilitate the nucleation of one phase with respect to the other during the transition, with a consequent suppression of $\Delta T_\text{hys}$, but without any change on the nucleation of the magnetic domains and only a small perturbation of their mobility.
%
In fact, contrary to the magnetic hysteresis, the other structural and magnetic properties of the film, are only slightly affected by the ion bombardment.
In particular, the large refrigeration power of MnAs related to GMCE is preserved.
This finding opens new perspectives on magnetic refrigeration  considering even bulk materials if dealing with defects that can be induced by highly charged ions at higher velocity taking advantage of their ballistic properties.\cite{Avasthi-Mehta}

We would like to thank  P.~Atkinson, M.~Barturen, M. Chatelet, R.~Gohier, J.~Mérot, D.~Mosca and L.~Thevenard for the fruitful discussion and  support.
This experiment is supported by a grant from ``Agence Nationale pour la Recherche (ANR)'' number \emph{ANR-06-BLAN-0223} and from Helmholtz Alliance HA216/EMMI.

%

\end{document}